\documentclass[conference]{IEEEtran}
\IEEEoverridecommandlockouts
\usepackage{cite}
\usepackage{amsmath,amssymb,amsfonts}
\usepackage{algorithmic}
\usepackage{graphicx}
\usepackage{textcomp}
\usepackage{xcolor}
\usepackage[raggedright]{sidecap}
\usepackage[singlelinecheck=false,justification=justified]{caption}
\usepackage{booktabs}
\usepackage{multirow}
\usepackage{xcolor}
\usepackage{bm}
\usepackage{caption}
\usepackage{subcaption}
\usepackage{sidecap}
\sidecaptionvpos{figure}{t}
\usepackage[colorlinks]{hyperref}

\def\BibTeX{{\rm B\kern-.05em{\sc i\kern-.025em b}\kern-.08em
    T\kern-.1667em\lower.7ex\hbox{E}\kern-.125emX}}
\begin{document}

\title{Frame-to-Utterance Convergence: A Spectra-Temporal Approach for Unified Spoofing Detection}

\author{Awais Khan$^{1}$, Khalid Mahmood Malik$^{1}$,Shah Nawaz$^{2}$\\
$^{1}$Department of Computer Science and Engineering, Oakland University, Rochester, Michigan, USA, \\
$^{2}$IMEC, Leuven, Belgium \\
}

\maketitle

\begin{abstract}
Voice spoofing attacks pose a significant threat to automated speaker verification systems. Existing anti-spoofing methods often simulate specific attack types, such as synthetic or replay attacks. However, in real-world scenarios, the countermeasures are unaware of the generation schema of the attack, necessitating a unified solution.
Current unified solutions struggle to detect spoofing artefacts, especially with recent spoofing mechanisms. For instance, the spoofing algorithms inject spectral or temporal anomalies, which are challenging to identify.
To this end, we present a spectra-temporal fusion leveraging frame-level and utterance-level coefficients. We introduce a novel local spectral deviation coefficient (SDC) for frame-level inconsistencies and employ a bi-LSTM-based network for sequential temporal coefficients (STC), which capture utterance-level artifacts. Our spectra-temporal fusion strategy combines these coefficients, and an auto-encoder generates spectra-temporal deviated coefficients (STDC) to enhance robustness. Our proposed approach addresses multiple spoofing categories, including synthetic, replay, and partial deepfake attacks. 
Extensive evaluation on diverse datasets (ASVspoof2019, ASVspoof2021, VSDC, partial spoofs, and in-the-wild deepfakes) demonstrated its robustness for a wide range of voice applications.
\end{abstract}


\section{Introduction}
\begin{figure*}[!t]
  \centering
  \centerline{\includegraphics[width=0.82\textwidth]{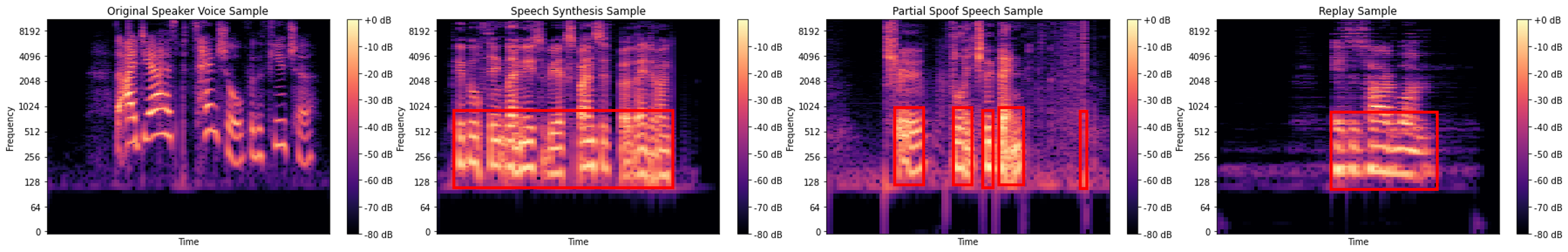}}
  \caption{Spectrogram comparison of bona fide (first-left), fully synthesized (second), partially deep fake (third), and replay (fourth) speech samples.}
  \label{fig:spect_analysis}
\end{figure*}

Voice authentication methods are mainstream solutions for identity verification systems, but the increasing prevalence of voice spoofing, including logical, physical, and deepfake attacks, poses a significant threat to their effectiveness~\cite{khan2023battling}. 
Existing methods often focus on mitigating individuals or a subset of these attacks, leaving systems vulnerable to others.
For example, a recent study~\cite{zhang2021initial} shows the limitations of existing systems, especially in detecting partial and full deepfake attacks. Notably, while existing systems successfully identify replay and synthetically fake speech samples~\cite{nautsch2021asvspoof,lavrentyeva2019stc,wang2021comparative}, they lack the ability to detect partial deepfake samples, as indicated in Table \ref{tab:crossperformance}.
These results show that the best-performing countermeasure in the ASVspoof$2019$ challenge demonstrates a substantial drop in performance when evaluated on partially spoofed samples. In another experiment, training the same system on the partial-spoof dataset and evaluating its performance on the ASVspoof$2019$ dataset yielded an intriguing result: performance on the development dataset remained competitive, but for the evaluation subset, the performance notably deteriorated.
As shown in Fig.~\ref{fig:spect_analysis}, unlike a spectrogram of replay or synthetic speech samples, a partial spoofing spectrogram indicates heterogeneous spectral artifacts. This may lead to significant performance deterioration of existing spoofing detection mechanisms, including utterance-level~\cite{alzantot2019deep,todisco2019asvspoof,lai2019assert}, transformer-based \cite{khan2023spotnet, ulutas2023deepfake} or deep learning~\cite{lavrentyeva2019stc,zeinali2019detecting,li2019anti,aljasem2021secure}. 
This is even true when these systems are trained on a specific dataset of partial spoofs, indicating the need for a solution adept at detecting frame-level disparities in partial deepfake scenarios.
\begin{table}[t]
\caption{ The best architecture from the ASVspoof challenge~\cite{zhang2021initial} is evaluated in terms of generalizability using ASVspoof$2019$-LA and Partialspoof$2021$ datasets. (Lower is better). The symbol $\textcolor{red}{\bm{\uparrow}}$ signifies the performance changes.}
\label{tab:crossperformance}
\centering
\resizebox{0.85\linewidth}{!}{
\begin{tabular}{|ll|ll|ll|} 
\hline
  &  & \multicolumn{2}{c}{ASV} & \multicolumn{2}{c|}{PSF} \\ \hline
                      &  Train & Dev. & Eval. & Dev. & Eval.  \\ \hline 
\multirow{2}{1.5cm}{EER(\%)} & ASV & 0.21 & 2.65 & 9.59 $\textcolor{red}{\bm{\uparrow}}$  & 15.96  $\textcolor{red}{\bm{\uparrow}}$    \\
  & PSF & 4.28 $\textcolor{red}{\bm{\uparrow}}$   & 5.38 $\textcolor{red}{\bm{\uparrow}}$  & 3.68   & 6.19 $\textcolor{red}{\bm{\uparrow}}$     \\ \hline
\multirow{2}{1.5cm}{min-tDCF}  & ASV  & 0.006 & 0.064  & 0.185  $\textcolor{red}{\bm{\uparrow}}$  & 0.300 $\textcolor{red}{\bm{\uparrow}}$   \\
& PSF & 0.115  & 0.171 & 0.100   & 0.164    \\
  \hline
\end{tabular}}
\end{table}

Earlier anti-spoofing methods were developed to prevent either physical or logical~\cite{khan2023battling,zhu2023local,wang2021comparative}. However, more recent methods are focused on creating a unified solution based on utterance-level features capable of detecting physical and logical attacks (LA)~\cite{aljasem2021secure, lai2019assert, lavrentyeva2019stc, zeinali2019detecting}. 
These unified solutions tend to be biased in favor of either detecting logical or physical attacks (PA). 
Thus, there is a need for an unbiased unified solution.
Moreover, other work has explored the detection of partial and fully deepfake attacks in a unified solution based on segment-level features~\cite{zhang2021multi, zhang2022partialspoof,zhu2023local}, however, these methods often fail to identify physical attacks.
Thus, full and partial deepfake attacks necessitate a comprehensive approach that accounts for both segment-level and utterance-level artifacts.
To this end, we propose a spectra-temporal approach that involves extracting frame-oriented spectral deviated coefficients (SDC), along with utterance-oriented sequential temporal coefficients (STC), using a Bidirectional Long Short-Term Memory (Bi-LSTM) network. These components collectively capture intricate patterns at both the utterance and frame levels, forming a strong foundation for our unified approach. In particular, the presented methodology unifies the detection of physical, partial, and fully deepfake attacks, resulting in a robust voice spoofing detection method.\\
The main contributions of this paper are as follows: 1) We introduce a spectra-temporal-based unified method for the detection of different voice spoofing categories. 2) We proposed spectral deviated coefficients for segment-level artifact extraction and employed a bi-LSTM network to capture sequential temporal artifacts within speech signals. Through rigorous experimentation using diverse datasets, we demonstrate the effectiveness of the proposed method. To the best of our knowledge, this is the first ever attempt to tackle four different types of voice spoofing with a single system.
\section{Proposed Method}
\begin{figure*}
  \centering
  \centerline{\includegraphics[width=0.82\textwidth]{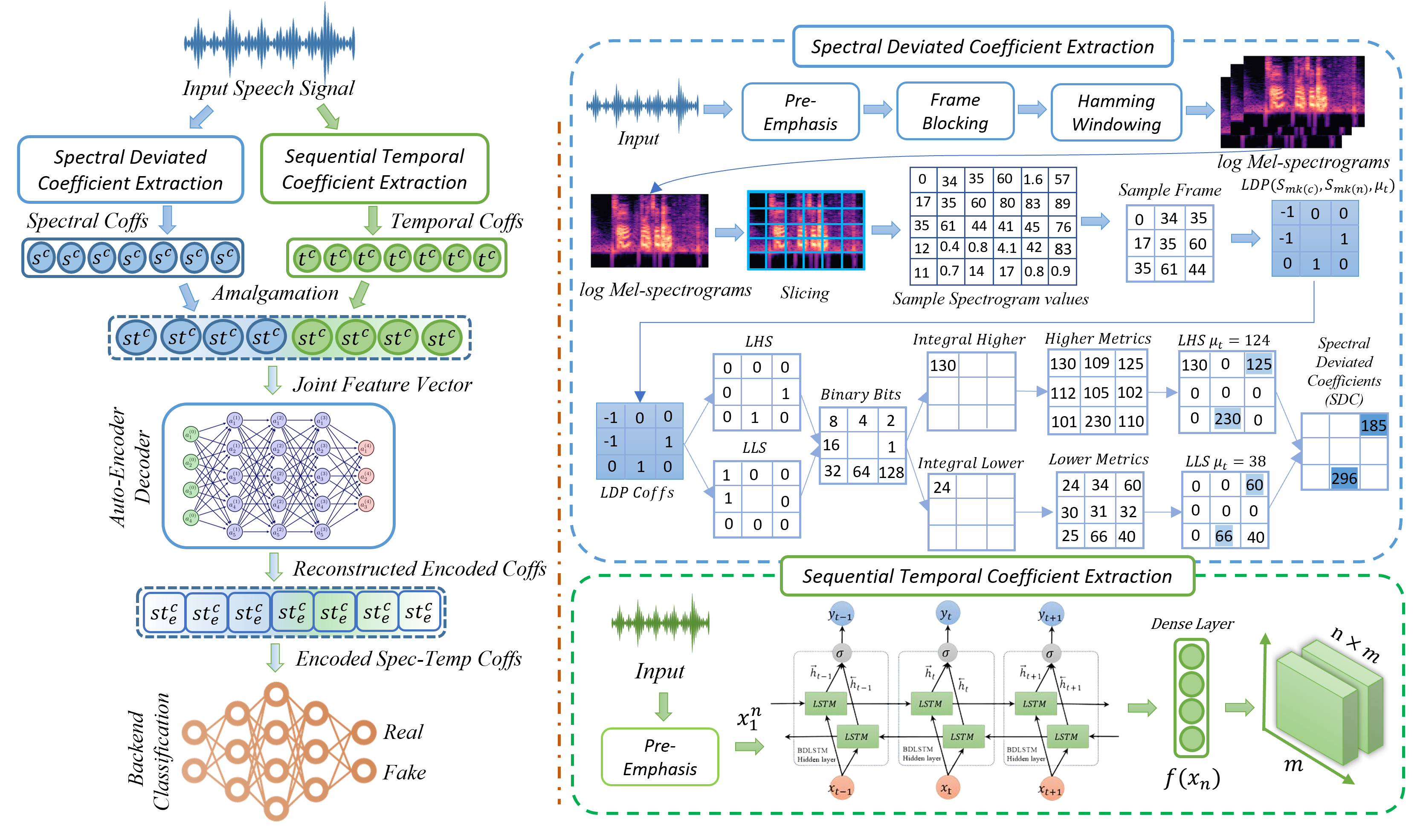}}
  \caption{Architectural diagram of the proposed solution (left). The right upper subset in blue dotted line represent the extraction mechanism of frame level Spectral Deviated Coefficients. The right lower subset in green presents the extraction mechanism of utterance level Sequential Temporal Coefficients. }
  \label{fig:main_arch}
\end{figure*}
The proposed method is divided into three sections, as shown in Fig.~\ref{fig:main_arch}. It consists of Spectral Deviated Coefficients (SDC), Sequential Temporal Coefficients (STC), and Spectra-Temporal Deviation Coefficients (STDC).
These sections collectively form a unified method for the reliable detection of voice spoofing.
\label{sec:pagestyle}
 \subsubsection{Spectral Deviated Coefficients (SDC)}
\label{sssec:subsubhead1} 
We used the raw input speech signal $s(t)$ to extract SDC, consisting of both higher and lower frequencies across various time frames:
\begin{equation}  s(t) = h * sin(2\pi f_1t) + l * sin(2\pi f_2t) \end{equation}
where $h$ and $l$ represent the amplitudes of frequencies, and $f_1$ and $f_2$ denote the higher and lower frequencies, respectively. Next, we use Hamming windows, which minimizes the spectral leakage by tapering frame edges and preventing abrupt truncation:
\begin{equation} w[n] = \alpha - \beta \cdot \cos \left(\frac{2 \pi n}{N - 1}\right) \end{equation}
\begin{equation}     y[n] = s[t] \cdot w[n] \end{equation}
where $s[t]$ denotes the input signal, $w[n]$ represents the Hamming window with a size of $N$, and $\alpha$ and $\beta$ are the window center and edge coefficients, respectively. The resulting segmented signal, after applying windowing and framing, is denoted as $y[n]$. Next, we transform the obtained $y[n]$ to the frequency spectra using a log-Mel spectrogram and fast Fourier transform (FFT) with the following parameters (hop length = $512$, mels = $128$, fft = $2048$) as follows:
\begin{equation} S[mk] = \log \left(1 + \sum_{n=0}^{N-1} |X[n]|^2 \cdot H_m[k,f_n] \right) \end{equation}
where \(S[mk]\) represents the log-Mel spectrogram at Mel frequency \(m\) and frame \(k\), \(X[n]\) stands for the Short-Time Fourier Transform (STFT) at time \(n\), and \(H_{m}[k,f_n]\) represents the Mel filterbank at frequency \(f_n\) corresponding to Mel frequency \(m\). The obtained log-transformed Mel spectrogram is then subjected to the Local Deviated Pattern (LDP) operator, which captures the local higher and lower frequency spectrum as follows:
\begin{equation}
  LDP(S_{mk(c)},S_{mk(n)},\mu_{t}) = \begin{cases}
    1 &  S_{mk(n)} \geq S_{mk(c)} + \mu_{t}, \\
    -1 &  S_{mk(n)} \leq S_{mk(c)} - \mu_{t}, \\
    0 & S_{mk(c)} - \mu_{t} \leq S_{mk(n)}, \\
    0 & S_{mk(n)} \leq S_{mk(c)} + \mu_{t}
\end{cases} \end{equation}
where $LDP(S_{mk(c)},S_{mk(n)},\mu_{t})$ represents the Local Deviated Pattern at position $(c,n)$ with $S_{mk(c)}$ and $S_{mk(n)}$ representing the central and neighboring window values, and \(\mu_t\) refers to the central tendency average of the window. We determine the conditioning threshold by considering both $S_{mk(c)}$ and $\mu_{t}$, rather than relying solely on the central window value. It enhances the extraction of LDP features by capturing deviations from the central value, revealing patterns indicative of underlying acoustic traits.

To efficiently handle spatial frequencies, we separately process the higher and lower frequencies of $S[mk]$. The LDP employs triplicate conditions to extract both higher and lower patterns. These patterns are further categorized into two sets: local higher spectra (LHS) and local lower spectra (LLS). Before computing $LHS$ and $LLS$, we transform negative values into positive ones, as shown in Eqs.~\ref{eq:lhs} and \ref{eq:lls}. For $LHS$, we convert all `-1' values to `0' while leaving the other values unchanged, as described in Eq.~\ref{eq:lhs} This results in a set of positive higher-order patterns in $S[mk]$. Similarly, LLS patterns are derived by replacing `1' with `0' and `-1' with `1' in $LDP(S_mk(c),S_mk(n),mu_t)$ as follows:
\begin{equation}
\label{eq:lhs}
    LHS= {LDP(S_{mk(c)},S_{mk(n)},\mu_{t}) =-1 \xrightarrow{} 0 }
\end{equation}  
\begin{equation}
\label{eq:lls}
    LLS= \begin{cases}
    {LDP(S_{mk(c)},S_{mk(n)},\mu_{t}) =1 \xrightarrow{} 0 } \\
    
    {LDP(S_{mk(c)},S_{mk(n)},\mu_{t}) =-1 \xrightarrow{} 1 }
    \end{cases}
\end{equation}
The binary bit streams, denoted as LHS and LLS, are converted into decimal values through a bit extraction process. We begin by extracting bits from the eastern direction and proceed in a counter-clockwise manner to obtain the decimal equivalents as shown in the equation below:
\begin{equation}    
    {HL}_{(int)} = \sum_{i=0}^{K-1} {HL}(C_{rn}) \times 2^{i-1}
\end{equation}
where ${HL}$ denotes the higher and lower coefficients obtained from Eqs.~\ref{eq:lhs} and \ref{eq:lls}, $C_{rn}$ represents the right neighbour at each position, and $K$ is the total number of bits. Next, we extract deviated tendency patterns from the obtained ${HL}_{(int)}$ to ensure the presence of spectral artifacts in both lower and higher spectral coefficients. Later, we only extract coefficients that exist in both higher and lower coefficients and neglect the rest of the values. We perform this task in a two-step process. First, we compute the mean vector of both higher and lower integrals separately as follows:
\begin{equation}
\label{eq:MV}
    MV_{(\delta)} = \frac{1}{n} \sum_{i=1}^{n} {HL}_{(int)}
\end{equation}
where $MV_{(\delta)}$ refers to the mean vector from higher and lower integrals ${HL}_{(int)}$. Next, we compute the central tendency vector from the obtained mean vectors $HT_{(\delta)}$ as follows:
\begin{equation}
    CTV_{(\delta)} = \frac{1}{n} \sum_{i=1}^{n} MV_{(\delta)}
\end{equation}
where $CTV_{(\delta)}$ denotes the central tendency mean value from the obtained mean vectors in Eq.~\ref{eq:MV}. By calculating the mean from the mean vectors, we confirm the presence of higher frequencies in both higher and lower integrals, combining them into a single optimal SDC. We retained values that are higher than their mean values and added them to derive the optimal robust spectral features, as demonstrated in Eq.~\ref{eq:SDC}.
\begin{equation}
\label{eq:SDC}
    SDC_{(coff)} = [{HL}_{(int)}>CTV_{(\delta)} ]
\end{equation}
where $SDC_{(coff)}$ represents the spectral deviated coefficients. Finally, a discrete Fourier transform (DFT) is applied to the LDP-transformed $SDC_{(coff)}$ coefficients to obtain robust 128D spectral features. The upper right side of Fig.~\ref{fig:main_arch} shows the extraction of SDC patterns.
\subsubsection{Sequential Temporal Coefficients (STC)}
\label{sssec:subsubhead2}
We employed a bidirectional long-short-term memory (Bi-LSTM) network to extract sequence-based utterance-level features. Bi-LSTM's bidirectional processing, unlike traditional LSTMs, considers both backward and forward context, enhancing complex temporal relationships.
In this work, a two-layer Bi-LSTM configuration was employed to improve temporal feature extraction, yielding $128$-dimensional temporal features.
\subsubsection{Spectra-Temporal Deviation Coefficients (STDC)}
\label{sssec:subsubhead3}
In this section, we focus on converging SDC and STC to create the Spectra-Temporal Deviation Coefficients (STDC) feature set. Given the distinct natures of SDC and STC, we address the range disparity by applying a tailored normalization technique that ensures both sets of coefficients are within a compatible range. The normalized coefficients are then processed through an autoencoder-decoder network, which distils the robust representation of spectra-temporal cues. The reconstruction process of the STDC feature set also aids in alleviating the challenges posed by sparsity in STC features before normalization.
\section{Experimentation and Results}
\label{sec:typestyle}
\subsection{Dataset and Implementation Details}
\label{sssec:subsubhead4}
We used several challenging datasets (ASVspoof$2019$~\cite{todisco2019asvspoof}, ASVspoof$2021$~\cite{yamagishi2021asvspoof}, VSDC~\cite{baumann2021voice}, partial spoofs~\cite{zhang2021initial} (Utterance-based), and in-the-wild audio deepfakes (IWA)~\cite{muller2022does}) to evaluate the proposed method. We used EER and accuracy to evaluate and compare the performance of the proposed method. We performed experiments on four NVIDIA Tesla V$100$ $16$G GPUs, coupled with $192$ GB of RAM and $48$ CPU cores operating at a clock speed of $2.10$ GHz. To address the data imbalance in ASVspoof$2019$ and partial spoof datasets, we applied five augmentation techniques from~\cite{COHEN202256,tak2022automatic}: high-pass filtering, low-pass filtering, compression, time and pitch shift, and reverberation. For our backend classifiers, we used a batch size of $32$, the Adam optimizer with an initial learning rate of $1e^{-4}$ and a weight decay of $0.001$. Models were trained for $50$ epochs using cross-entropy loss.
\subsection{Experimental Results}
\label{sssec:subsubhead5}
\subsubsection{Performance Analysis of the SDC with Different Classifiers}
\label{sssec:subsuhead6}
We have evaluated the performance of the proposed SDC features with different machine learning (ML) and residual-based classifiers, and the results are presented in Table~\ref{tab:alone_sdc}. It is observed from the results that SDC features performed well with both ML and residual classifiers, with the best performance achieved with Ensemble and SE-ResNext$18$ classifiers. The lower EERs show the efficiency of the presented coefficients and their potential standalone use for voice spoofing attack detection.
\subsubsection{Performance Analysis of STDC with Different Voice Spoofing Datasets}
\label{sssec:subsubhead7} 
We choose the best-performing back-end classifier (SE-ResNeXt18) from Table~\ref{tab:alone_sdc} and evaluate the performance of the proposed system with different datasets.
Results are shown in Table~\ref{tab:STDC}, indicating performance improvement when spectral coefficients converge with temporal coefficients. Specifically, EER improves from $0.25$ to $0.22$, $0.60$ to $0.52$, $3.70$ to $3.50$, and so on.
These results show the significance of incorporating both spectral and temporal coefficients.
\begin{table}[b]
\caption{\small Performance analysis of spectral deviated coefficients with different ML and residual-based back-end classifiers (lower is better).}
\centering
\resizebox{0.89\linewidth}{!}{
\begin{tabular}{|l|c|c|c|c|c|c|c|}
\hline
Classifiers & \multicolumn{2}{|c|}{ASV-19} & \multicolumn{2}{|c|}{ASV-21} & PSF & VSDC & IWA\\
\hline
   & LA & PA & LA & DF &  &  &  \\
\hline
        Random Forest & 0.49 & 1.20 & 4.37 & 5.11 & 7.90 & 2.30 & 2.90  \\
                 KNN  & 0.28 & 1.00 & 4.17 & 5.20 & 6.11 & 1.89 & 2.51  \\
                 SVM  & 0.22 & 0.70 & 4.90 & 3.95 & 5.95 & 1.02 & 0.70  \\
 Logistics Regression & 0.30 & 0.80 & 3.95 & 3.90 & 6.30 & 2.15 & 0.90  \\
         Naive Bayes  & 0.31 & 0.75 & 4.98 & 4.50 & 6.50 & 2.40 & 0.95  \\
        Decision Tree & 0.45 & 0.90 & 5.30 & 5.50 & 7.11 & 3.90 & 1.90  \\
             Ensemble & 0.26 & 0.63 & 3.79 & 3.40 & 6.02 & 2.01 & 0.40   \\
             ResNet18 & 0.28 & 0.60 & 4.01 & 3.30 & 5.95 & 1.56 & 0.45   \\
          SE-ResNet18 & 0.29 & 0.63 & 3.90 & 3.40 & 5.98 & 1.10 & 0.35   \\
            ResNext18 & 0.25 & 0.65 & 3.98 & 3.35 & 6.00 & 1.50 & 0.40  \\
SE-ResNext18 & \textbf{0.25} & \textbf{0.60} & \textbf{3.70} & \textbf{3.41} & \textbf{5.98} & \textbf{0.95} & \textbf{0.32} \\        \hline
\end{tabular}}
\hspace{0.08em}
\label{tab:alone_sdc}
\end{table}
\begin{table}[!htp]
\caption{\small Performance analysis of Spectra-Temporal Deviated Coefficients against different datasets (Lower is better).}
\centering
\resizebox{0.80\linewidth}{!}{
\begin{tabular}{|l|c|c|c|c|c|c|c|}
\hline
Performance & \multicolumn{2}{|c|}{ASV-19} & \multicolumn{2}{|c|}{ASV-21} & PSF & VSDC & IWA\\
\hline
   & LA & PA & LA & DF &  &  &  \\
\hline
         EER  & 0.22 & 0.52 & 3.50 & 3.20 & 5.90 & 0.80 & 0.30  \\
           Accuracy (\%)  & 98.5 & 98.0 & 95.5 & 95.0 & 93.5 & 98.5 & 98.5  \\ \hline
\end{tabular}}
\hspace{0.08em}
\label{tab:STDC}
\end{table}
\subsubsection{Comparison with Existing Methods}
\label{sssec:subsubhead8}
We evaluate our proposed methods against recent voice spoofing countermeasures, addressing four distinct attack types: LA, PA, and fully and partially deepfake. To our knowledge, this is one of the first comprehensive approach to tackle these four attack categories simultaneously.
Moreover, we compared our solution to specific attack-focused methods, such as ASVspoof$2019$ (LA+PA) in Table~\ref{tab:STDC-ASV19}, ASVspoof$2021$ in Table~\ref{tab:STDC-ASV21}, partial-spoof in Table~\ref{tab:STDC-PSF}, and IWA in Table~\ref{tab:STDC-IWA}.
Our method outperforms existing state-of-the-art methods. Though the performance of the method on specific dataset (IWA) and some attacks (PSF) is slightly higher, it exhibits superior generalizability across a wide range of attacks, providing a holistic defense mechanism with enhanced detection capabilities.
\begin{SCtable}[][!htp]
\caption{\small Comparison of proposed method with existing methods on ASVspoof2019 dataset (Lower is better).}
\centering
\resizebox{0.50\linewidth}{!}{
\begin{tabular}{|l|c|c|c|}
\hline
Study & Method & \multicolumn{2}{|c|}{ASV-19} \\
\hline
   &  &   LA & PA \\ \hline
         \cite{sahidullah2023introduction}  & CQCC-GMM & 9.87 & 11.04   \\
          \cite{sahidullah2023introduction}  & LFCC-GMM & 11.96 & 13.54  \\ 
        \cite{rupesh2021novel}  & FBCC-GMM & 6.16 & 10.36  \\
          \cite{li2021replay}  & SE-Res2Net50 & 2.86 & 1.00   \\ 
        \cite{monteiro2020generalized} & LFCC-CNN & 9.09 & 2.01  \\
          \cite{lavrentyeva2019stc} & CQT-DCT-LCNN & 1.84 & 0.54   \\      
          \cite{lai2019assert} & ASSERT & 6.70 & 0.59   \\
           \cite{ren2023generalized}  & Knowledge Amalgamation & 2.39 & 1.97   \\ 
           \textbf{ Ours } & \textbf{STDC+SE-ResNeXt18} & \textbf{0.22} & \textbf{0.52}   \\    
\hline
\end{tabular}
}
\hspace{0.08em}
\label{tab:STDC-ASV19}
\end{SCtable}
\begin{SCtable}[][!htp]
\caption{\small Comparison of proposed method with existing methods on ASVSpoof2021 dataset (Lower is better). }
\centering
\resizebox{0.50\linewidth}{!}{
\begin{tabular}{|l|c|c|c|}
\hline
Study & Method & \multicolumn{2}{|c|}{ASV-21} \\
\hline
   &  &   LA & DF \\ \hline
         \cite{tak2022automatic}  & wav2vec 2.0 & 1.19 & 4.38   \\
          \cite{tomilov21_asvspoof}  & LCNN+ResNet+RawNet & 1.32 & 15.64  \\ 
        \cite{das2021known}  & GMM+LCNN (Ensemble) & 3.62 & 18.30  \\
           \cite{chen2021ur} & ECAPA-TDNN (Ensemble) & 5.46 & 20.33   \\ 
        \cite{chen2021pindrop} & ResNet (Ensemble) & 3.21 & 16.05  \\
           \cite{wang2021investigating}  & W2V2 (fixed)+LCNN+BLSTM & 10.97& 7.14   \\      
           \cite{wang2021investigating}  & W2V2 (finetuned)+LCNN+BLSTM & 7.18 & 5.44   \\
           \cite{martin2022vicomtech} & FIR-WB & 3.54 & 4.98   \\ 
            \textbf{Ours}  & \textbf{STDC+SE-ResNeXt18} & \textbf{3.50} & \textbf{3.20}  \\    
\hline
\end{tabular}
}
\hspace{0.10em}
\label{tab:STDC-ASV21}
\end{SCtable}

\begin{SCtable}[][!htp]
\caption{\small Comparison of proposed method with existing models on partial spoof dataset (Lower is better).}
\centering
\resizebox{0.50\linewidth}{!}{
\begin{tabular}{|l|c|c|}
\hline
Study & Method & EER \\
\hline
         \cite{zhu2023local}  & LCNN & 6.19 \\
          \cite{zhu2023local}  & SELCNN & 6.33   \\ 
        \cite{zhu2023local}  & H-MIL (Ensemble) & 5.96  \\
           \cite{zhu2023local} & LS-H-MIL & 5.89   \\ 
        \cite{zhang2021multi} & LCNN + LSTM & 8.61   \\
           \cite{zhang2021multi}  & SELCNN(2)+LSTM & 7.69 \\      
            \textbf{Ours } & \textbf{STDC+SE-ResNeXt18} & \textbf{5.90}  \\    
\hline
\end{tabular}
}
\hspace{0.08em}
\label{tab:STDC-PSF}
\end{SCtable}
\begin{SCtable}[][!htp]
\caption{\small Comparison of proposed method with existing models on In-the-Wild Audio deepfake dataset (Lower is better). }
\centering
\resizebox{0.50\linewidth}{!}{
\begin{tabular}{|l|c|c|}
\hline
Study & Method & EER \\
\hline
         \cite{wang2021comparative}  & LCNN+LSTM & 0.33 \\
          \cite{tak2021end}  & RawGAT & 0.53  \\ 
        \cite{tak2021end}  & RawNet2 & 0.51  \\
           \cite{desplanques2020ecapa} & ECAPA-TDNN & 0.30   \\ 
        \cite{chung2020defence} & H/ASP & 0.27  \\
           \cite{heo2020clova}  & ClovaAI & 0.36 \\      
            \textbf{Ours}  & \textbf{STDC+SE-ResNeXt18} & \textbf{0.30 } \\    
\hline
\end{tabular}
}
\hspace{0.08em}
\label{tab:STDC-IWA}
\end{SCtable}
\section{Conclusion and Future work}
\label{sec:typestyle2}
We have presented a spectra-temporal approach for the detection of a wide range of voice spoofing attacks. Our proposed method incorporates spectral deviated coefficients, sequential temporal coefficients, and spectra-temporal deviated coefficients obtained through segment-level and utterance-level patterns. Our method successfully addresses various voice spoofing attacks, such as logical, physical, full, and partial deepfake attacks, within a unified framework, marking it as a significant advancement in the field of voice spoofing detection. The effectiveness of our proposed method has been rigorously evaluated against state-of-the-art unified classifiers, highlighting its potential to enhance voice spoofing detection across a wide range of voice attack scenarios.
\clearpage
\bibliographystyle{IEEEbib}
\bibliography{IEEEbib}

\end{document}